# THE STABILITY OF AN OPTICAL CLOCK LASER TRANSFERRED TO THE INTERROGATION OSCILLATOR FOR A CS FOUNTAIN

B. Lipphardt, G. Grosche, U. Sterr, Chr. Tamm, S. Weyers and H. Schnatz

*Abstract*— We stabilise a microwave oscillator at 9.6 GHz to an optical clock laser at 344 THz by using a fibre-based femtosecond laser frequency comb as a transfer oscillator. With a second frequency comb we measure independently the instability of the microwave source with respect to another optical clock laser frequency at 456 THz. The total fractional frequency instability of this optic-to-microwave and microwave-to-optic conversion resulted in an Allan deviation $\sigma_y$, of $\sigma_y = 1.2 \times 10^{-14}$ at 1s averaging time (band width 50 kHz). The residual phase noise density is -97 dBc / Hz at 10 Hz offset from the 9.6 GHz carrier. Replacing the existing quartz-based interrogation oscillator of the PTB caesium fountain CSF1 with this optically stabilised microwave source will reduce the instability contribution due to the Dick effect from the $10^{-13}$-level at 1s averaging time to an insignificant level at the current status of CSF1. Therefore this new microwave source can be an alternative to cryogenic sapphire-loaded cavity oscillators in order to overcome the limitations of state-of-the-art quartz oscillators.

*Index Terms*— fountain clocks, optical frequency standards, frequency combs, microwave oscillators, phase noise measurement

I. INTRODUCTION

In existing atomic caesium fountains used for the realization of the SI unit "second", an intrinsic frequency instability contribution is caused by the phase noise of the employed microwave source, an effect often called "Dick effect" [1, 2]. To overcome this limitation, typically at the low $10^{-13}$- level at 1s averaging time, an ultra-stable cryogenic sapphire-loaded-cavity-oscillator [3] has been used, and quantum-projection- noise limited operation down to the low $10^{-14}$ - level has been demonstrated [4, 5]. Here we present an alternative approach using an optically stabilized microwave source designed to replace the quartz based interrogation oscillator of the PTB caesium fountain CSF1. The resulting instability is expected to be dominated by quantum projection noise only.

As ultra-stable clock lasers have become new sources of superior stability [6, 7] and femtosecond combs can transfer this stability to the microwave domain [8], these two key elements allow a novel way of microwave generation based on optical clock lasers. While this combination is routinely used for absolute frequency measurements [9, 10], it also has an enormous potential for generating ultra low-jitter microwave signals in a novel way.

In 2005, Bartels *et al.* showed that in a synthesized 10 GHz signal the femtosecond comb added an instability of $\sigma_y = 6.5 \times 10^{-16}$ at 1 s [11]. For averaging times longer than 1 s the stability suffered from $1/f$ frequency noise contribution of the comb. At present, the best reported residual phase noise from such a system is approximately $(-95\,\text{dBc}/\text{Hz})/f^2$ for Fourier frequencies $1\,\text{Hz} < f < 1\,\text{kHz}$ approaching a white noise level of $-140\,\text{dBc}/\text{Hz}$ at Fourier frequencies above 10 kHz offset from the 10 GHz carrier [12]. The studies described above, using two femtosecond combs stabilized to a common optical reference, permit tests of the fidelity of the frequency division process from optical to microwave frequencies. Alternatively, using two independent combs stabilized to two independent optical references provides information on the absolute stability, reproducibility and frequency accuracy. With this latter configuration, Bartels *et al.* [11] have demonstrated the synthesis of 10 GHz signals having a fractional frequency instability of less than $\sigma_y = 3.5 \times 10^{-15}$, limited by the stability of the poorer of the two optical references. Similar results have been obtained by Kim *et. al.* using a balanced optical–microwave phase detector for the extraction of microwave signals at 10 GHz from optical pulse trains [13].

These studies were all carried out using Ti:Sapphire-based femtosecond combs. However, for applications that require long-term continuous operation, fibre-based combs are a more attractive option, but there is little published work on the use of such systems for low-noise microwave synthesis.

In the only experiment published to date [14], the stability of a 10 GHz signal derived from a fibre comb was compared with an established high stability frequency comb based on a Ti:Sapphire laser. Both combs were locked to a common optical reference with a stability of around $\sigma_y = 3 \times 10^{-15}$ at 1 s. For the 10 GHz signal a short-term stability of $\sigma_y = 2 \times 10^{-14}$ at 0.1 s was achieved. These results already indicated that a microwave signal generated from an optical clock will be an excellent candidate to interrogate microwave fountain clocks.

In this study we present a novel way to synthesize an ultra low noise microwave frequency using an ultra-stable clock laser, a femtosecond comb that transfers this stability to the microwave domain and a low noise microwave source.

We use a frequency measuring system that routinely compares the frequency of the $Yb^+$ frequency standard of PTB [15] with that of a Cs fountain clock, CSF1 [16]. The ultra-stable $Yb^+$ clock laser is measured using a commercially available fibre-based femtosecond comb [17]. Its 100 MHz repetition rate $f_{rep}$ is locked with a few 100 Hz bandwidth to a reference frequency from a H-maser. The self referencing of the comb is arranged by the f-2f scheme - stabilising the carrier offset frequency $\nu_{ceo}$ is achieved with a bandwidth of several kHz.

As for fs-fibre combs the white frequency noise extends to Fourier frequencies up to 100 kHz, the limited bandwidth of the control elements puts some constraints on ultra-precise measurements. We circumvent this problem by using the transfer-oscillator concept derived by Telle et. al. [18]. In this case the femtosecond comb needs no fast servo control loops. More details are described in Ref. [19].

The optical frequency of a clock laser $\nu_L$ is derived from a simultaneous measurement of the three radio frequencies and the known mode number $m$ according to

$$\nu_L(t) = m \cdot f_{rep}(t) + \nu_{ceo}(t) + \nu_x(t) \quad (1),$$

where $\nu_x$ is the beat signal of the $Yb^+$ clock laser $\nu_L$ with the comb line nearest to it.

We upgraded our current frequency measurement system by adding a module that allows a simultaneous generation of a microwave-signal. Additionally, a determination of the phase noise of highly stable microwave oscillators in real time is easily accomplished.

II. SETUP

In [11] a microwave signal was directly generated in a photo diode, detecting a high harmonic of the pulse repetition rate. Instead, we start with a microwave oscillator, which is phase locked to an optical reference using the fs comb.

The microwave source is a commercial 9.6 GHz dielectric resonator oscillator (DRO) with a state-of-the-art phase noise of -115 dBc / Hz at 10 kHz from the carrier (see fig. 3). The frequency control input has a bandwidth of > 1 MHz.

The setup for locking the DRO to an optical standard is shown in Fig. 1. The DRO's output is mixed with the 96th harmonic of the comb's repetition frequency ($a \times f_{rep}$), resulting in an intermediate beat signal $\nu_d$ at approximately 6 MHz. This signal is subsequently multiplied by a factor $b$ using a harmonic tracking oscillator.

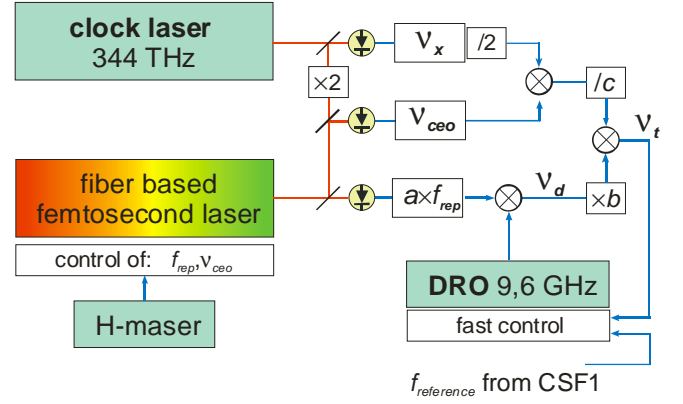

Fig. 1: Set-up for absolute frequency measurements and phase locking of a microwave source. DRO: dielectric resonator oscillator; a, b ∈ N, c ∈ R; For more details see text.

The up-conversion of the microwave signal is accompanied by a down-conversion of the optical signal. In a first step we take into account that the beat signal $\nu_x$ is derived from a beat with the frequency doubled output of the comb; therefore this signal has to be divided by 2 before it is mixed with the carrier offset frequency $\nu_{ceo}$. Subsequent division of the sum ($\nu_{ceo} + \nu_x/2$) by a factor $c$ allows one to generate a beat signal $\nu_t$ between the up-converted microwave signal and the down-converted optical beats at an intermediate virtual frequency. The divisor $c$ is chosen according to

$$m = a \times b \times c \quad (2).$$

This rational divider is realized by a direct digital synthesizer (DDS) with a resolution of 48 bit. With $\nu_d = \nu_{DRO} - a \times f_{rep}$ and $\nu_t = (\nu_{ceo} + \nu_x/2)/c - b \times \nu_d$ we obtain

$$\nu_t = \nu_L/c - b \, \nu_{DRO} . \quad (3)$$

This signal corresponds to a virtual beat between the $Yb^+$ clock laser and the DRO. In our case this signal is measured at an intermediate frequency of about 77 GHz.

Our choice of the divider ratios of $b$ and $c$ was guided by the following considerations: increasing the factor $b$ decreases the demands on the phase noise of the DDS (divider $c$) and the subsequent electronics while noise issues are passed to the multiplier $b$. In our case, the phase delay of the multiplier limits the attainable electronic locking bandwidth of the harmonic tracking filter and leads to an optimum factor of $b = 8$.

Phase locking of the DRO is achieved by comparing the beat frequency $\nu_t$ with a reference frequency $f_{reference}$. In order to lock the DRO to a Cs fountain clock for long integration times this reference frequency is steered by the atomic fountain clock.

This concept has two major advantages: it allows a comfortable, real time measurement of the phase noise of the DRO or a locking of the DRO to an optical clock with a bandwidth up to several MHz.

For a verification of the DRO's performance with respect to phase noise and instability a second independent measurement system is required. For this purpose, we use a second frequency comb in combination with the clock laser of the Ca optical frequency standard [20].

Both clock laser systems with their accompanying fs- combs are located in different buildings and are linked by 300 m of coaxial cable (type: FSJ1). In order to avoid degradation of the SNR we transmit the frequency of the DRO divided by a factor of 8.

At the Ca laser setup the transmitted frequency of 1.2 GHz is frequency doubled and compared with the 24[th] harmonic of the Ca- fs comb in the same way as described above. The difference in the nearly identical set-ups is that the optical frequency of the Ca system is at 456 THz and that the transfer signal $\nu_t$ (Ca) is only used for analysis.

In order to achieve the highest possible resolution the factor b here is set to 1024 and the divider c is adjusted accordingly. The phase noise of the DRO is thus analysed at a virtual frequency of 2.4 THz. Multiplied by such a huge factor, the noise of the DRO is measurable with a conventional spectrum analyser.

For further data analysis and processing in the time domain, we use a multichannel accumulating counter with synchronous readout and zero dead-time, referenced to a H-maser that is controlled by the caesium fountain clock CSF1.

III. RESULTS

The Yb[+] clock laser provides the stability in the optical domain (triangles in Fig. 2). The curve shown is the result of an optical-to-optical frequency comparison between the Ca- clock laser and the Yb[+]-clock laser. For very short times τ< 0.1 s the Allan deviation $\sigma_y(\tau)$ is dominated by the linewidth of the clock lasers. For intermediate times 0.2 s < τ < 50 s the stability is limited by the thermal noise of the optical reference resonators, while for τ > 50 s, $\sigma_y(\tau)$ increases due to the relative drift of the resonators [21].

For an ideal down conversion of this stability by means of a fs-comb we expect to achieve the same low instability for our microwave source. Using our frequency counting system, we measured the Allan deviation of the stabilized microwave signal. As shown in Fig. 2 (dots), the instability (measured with a high frequency cut-off of 50 kHz) exhibits a 1/τ behaviour for τ < 10 s, resulting in a $\sigma_y(1\ s) = 1.2 \times 10^{-14}$ at 1 s averaging time. At τ ~ 10 s, the stability reaches that of the optical frequency standard. For longer averaging times τ > 50 s the residual drift between the two optical resonators leads to an increase of $\sigma_y$. For comparison, we show the stability of the in-loop signal (squares), the stability of the 5 MHz quartz (open grey circles) currently used in a synthesis chain for the CSF1 fountain clock [22] and typical data of an ultra-stable cryogenic sapphire-loaded cavity-oscillator (asterisks).

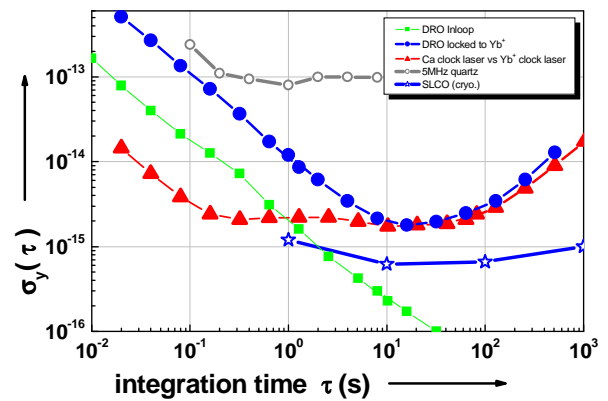

Fig. 2: Measured Allan deviation of the locked DRO (dots). The frequency comparison between optical clock lasers is shown as triangles and that of the in-loop signal as squares. For comparison we have included the stability of a 5 MHz quartz oscillator and that of an cryogenic sapphire-loaded cavity-oscillator.

In the following we discuss in the phase noise domain some technical limitations that hamper the ideal performance:

As we detect a high harmonic of the pulse repetition frequency with an ultra-fast photo detector, the achievable signal to noise ratio (SNR) is limited by shot noise and the dynamic range of the photo diode. In our case, this causes a white phase noise level of -134 dBc / Hz, which is indicated by the constant line in Fig. 3.

Another technical restriction is due to the fact that the generated microwave signal is transmitted over 300 m coaxial cable to the analysing system in another building. At 9.6 GHz attenuation of the cable is significant and would degrade the SNR. To avoid this, we divide the signal by 8. At 1.2 GHz we achieve a similar SNR for the transmitted signal and the signal derived from the photodiode of the analysing fs comb. The price to be paid is additional phase noise of the divider. The specification of the divider's phase noise is shown in Fig. 3 as diamonds. For Fourier frequencies f < 10 kHz, the noise of the divider is above the level due to the photo-diode and is dominated by $1/f$ phase noise, resulting in -110 dBc / Hz at an offset of $f$ = 10 Hz. While the free running DRO has excellent phase noise at Fourier frequencies above several MHz, its low frequency noise is dominated by flicker of frequency noise. This reaches -56 dBc / Hz at 100 Hz offset from the 9.6 GHz carrier.

The crossing point at 50 kHz of the DRO's phase noise with that of the photo diode's shot noise determines the ideal locking bandwidth. The PLL for phase-locking the DRO frequency to the optical frequency standard uses an additional second integrator at 12 kHz. This results in a significant reduction of the DRO's phase noise (red curve in Fig. 3). The total phase noise exhibits a white phase noise level of -123 dBc / Hz for 3 kHz <f <50 kHz. (Data beyond 50 kHz reflects a roll-off with $1/f^3$ due to a tracking oscillator of the analysing system.) This white phase noise level can be changed by adjusting the optical power at the photo diode. Similar observations have been reported by Newbury *et al.* [23] and point to a possible AM / FM conversion within the photo detector. This limitation will be studied in more detail in future.

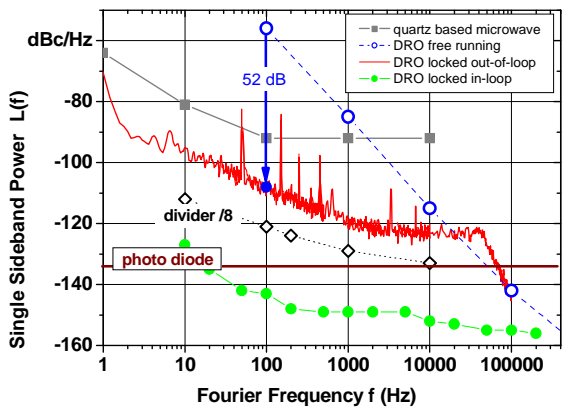

Fig. 3: Single sideband phase noise (PN) of a DRO at 9.6 GHz: free running PN (blue open circles), reduced noise, when the DRO is locked to an optical clock (red curve); the brown line indicates the current limitation due to the detection of $f_{rep}$. Diamonds show the noise specification of a frequency divider. For comparison the phase noise of a state-of-the-art 5 MHz quartz oscillator (grey squares) and that of the in-loop signal (green dots) are shown. The in-loop signal was recorded for a locking bandwidth of several hundred kHz.

For Fourier frequencies $f < 3$ kHz the noise is dominated by flicker of phase noise. This level of 1/f-noise is about 10 dB above the noise expected from the specifications of the frequency divider. At $f = 100$ Hz the phase noise is suppressed by 52 dB with respect to the noise of the free running DRO and reaches L(f)=-95 dBc / Hz at 10 Hz offset from the 9.6 GHz carrier.

The integrated phase noise up to a high frequency cut-off of 50 kHz leads to an Allan deviation of $\sigma_y(\tau) = 9 \times 10^{-15}/\tau$. This is in excellent agreement with the data derived from the time domain measurements (see Fig. 2).

The corresponding in-loop signal of the stabilized DRO, as analysed by the phase noise of the transfer beat $\nu_t$ (green dots in Fig. 3) is well below -140 dBc / Hz for $f > 40$ Hz, showing a slight increase for $f < 40$ Hz in good agreement with the Allan standard deviation of $\sigma_y(\tau) = 2 \times 10^{-15}/\tau$ of the in-loop signal (squares) shown in Fig. 2. Again for comparison, we additionally show the phase noise of a 5 MHz quartz oscillator mentioned above.

We have achieved a continuous operation of the complete setup over several days. We have thus obtained a reliable, self-contained module for synthesising an ultra-low noise microwave frequency.

## IV. Prospects: An Interrogation Oscillator For A Fountain Clock

The PTB caesium fountain CSF1 currently uses a recently developed new 9-GHz synthesis chain [22]. In this synthesis a 9.6 GHz YIG oscillator is locked to a 5 MHz quartz oscillator via a divider chain. The signal from the atoms is then used to steer the frequency of the quartz oscillator whose instability specification is shown in Fig. 2.

Mainly caused by the time needed for loading and detecting the atoms, the pulsed operation mode of a caesium fountain comprises a significant amount of dead time during which the quartz oscillator frequency is not controlled by the atomic resonance signal. Such a dead time results in a degradation of the fountain frequency instability caused by the frequency noise of the interrogation oscillator (Dick effect) [2].

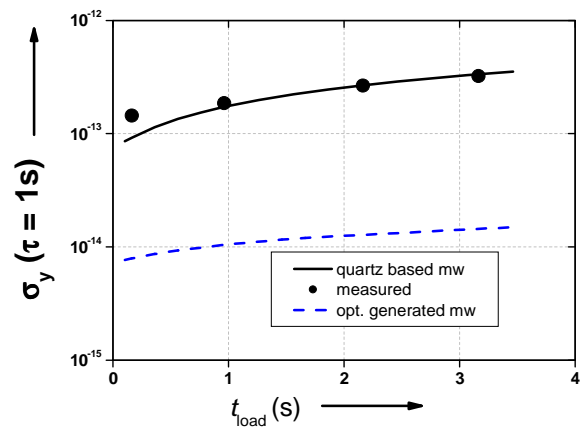

Fig. 4: Allan standard deviation for different loading (dead) times. Black symbols: measured frequency instability of CSF1 using the quartz oscillator based microwave source and a hydrogen maser as reference . Solid and dashed lines: calculated contributions due to the Dick effect using the quartz oscillator based or the optically stabilized microwave source, respectively.

For CSF1 long loading times of the magneto-optical trap are used for increasing the atom number and evaluating the collisional frequency shift [16]. In Fig. 4, the solid line shows the calculated instability contribution due to the Dick effect for CSF1, when the loading time and thus the dead time is varied. The calculation is based on the phase noise data from the data sheet of the employed quartz crystal oscillator and the sensitivity function calculated for CSF1 [2]. With increasing dead time, an increasing number of oscillator phase noise components at small Fourier frequencies contributes and thus degrades the stability. The black data points depict a set of measured frequency instabilities using a hydrogen maser as a reference for CSF1. For loading times longer than half a second the measured instability is clearly dominated by the local oscillator noise via the Dick effect.

If the instability contribution due to the Dick effect is calculated based on the measured phase noise (Fig. 3) of the optically stabilised microwave source, the data shown by the dashed line in Fig. 4 are obtained. At this level, which is at or even below $\sigma_y(1\ s) = 1 \times 10^{-14}$, the instability contribution due to the Dick effect would have a negligible effect on the overall CSF1 instability, which then would be quantum projection noise limited by the currently accessible numbers of detected atoms for different loading times. From these numbers it can be expected that for normal operation at short loading time the overall instability is reduced by a factor of 2 by employing the optically stabilized microwave source instead of the quartz based source. For the long loading times used in collisional shift evaluations, even larger improvements up to a factor of 4 can be achieved.

V. CONCLUSION

We have realized a highly stable microwave source at 9.6 GHz using a fibre-based fs-frequency-comb and an optical reference frequency. We demonstrated that the stability of the optical clock laser can be transferred to the microwave domain without tight locking of the frequency comb, generating a stabilized microwave signal with a stability superior to common 9 GHz synthesizers based on ultra-stable 5 MHz quartz oscillators.

With an achieved short term stability of $1 \times 10^{-14}$ at 1 s of the optically stabilized DRO, the Dick effect would give a negligible contribution to the overall CSF1 instability which then would be quantum projection noise limited. Further improvements can be achieved by increasing the number of atoms by loading from an atomic beam [5].

ACKNOWLEDGMENT

This work was supported in part by the German Science Foundation DFG through SFB407.